\documentclass[prb,showpacs,footinbib,twocolumn,final]{revtex4}
\usepackage{graphicx}
\usepackage{latexsym}
\usepackage{amsmath}
\usepackage{amssymb}
\usepackage{bm}
\usepackage{wasysym}
\renewcommand{\S}{{\cal S}}

\begin{document}
\title{Semiclassical analysis of edge state energies in the integer quantum Hall effect}
\author{Yshai Avishai$^{1-4}$ and Gilles Montambaux$^4$}

\affiliation{$^1$Department of Physics and Ilse Katz Center for
Nanotechnology, Ben Gurion University, Beer Sheva 84105, Israel}
\affiliation{$^2$RTRA -- Triangle de la Physique, Les Algorithmes,
91190 Saint-Aubin, France}
 \affiliation{$^3$Institut de Physique Th\'eorique,
CNRS URA-2306, CEA Saclay, 91191 Gif-sur-Yvette Cedex, France}
\affiliation{$^4$Laboratoire de Physique des Solides, CNRS UMR-8502
Universit\'e Paris Sud, 91405 Orsay Cedex, France}
%
%
\begin{abstract}
Analysis of edge-state energies in the integer quantum
Hall effect is carried out within the semiclassical  approximation.
When the system is wide so that each edge can be considered
separately, this problem is equivalent to that of a one dimensional
harmonic oscillator centered at $x=x_{c}$ and an infinite wall at
$x=0$, and appears in numerous physical contexts. The eigenvalues
$E_{n}(x_{c})$ for a given quantum number $n$ are solutions of the
equation $S(E,x_{c})=\pi [n+ \gamma(E,x_{c})]$ where $S$ is the WKB
action and $0<\gamma<1$ encodes all the information on the
connection procedure at the turning points.
 A careful implication of the WKB connection formulae results in an
excellent approximation to the exact energy eigenvalues. The
dependence of $\gamma [E_{n}(x_{c}),x_{c}] \equiv \gamma_{n}(x_{c})$
on $x_{c}$  is analyzed between its two extreme values $\frac{1}{2}$
as $x_{c} \to -\infty$ far inside the sample
and $\frac{3}{4}$ as $x_{c} \to \infty$ far outside the sample.
The edge-state energies $E_{n}(x_{c})$ obey an almost exact scaling
law of the form
$E_{n}(x_{c})=4[n+\gamma_{n}(x_{c})]f(\frac{x_{c}}{\sqrt{4n+3}})$
and the scaling function $f(y)$ is explicitly elucidated.
\end{abstract}
\date{june 18, 2008}
\maketitle
%

\section{Introduction and statement of the problem}
\label{intro} The concept of edge states is central for elucidating
the physics of quantum Hall and spin quantum Hall systems
\cite{BH,MD,MB,CK}. It is illustrated by considering an electron
(mass $m$ and charge $-e$) restricted in two dimensions to a stripe
$-\infty < y < \infty, \ \ \ -L_{x} \le x \le 0$, and acted upon by
a perpendicular magnetic field ${\bf B}=B {\hat {\bf z}}={\bm
\nabla} {\bf \times} {\bf A}$.  At strong enough magnetic field it
is reasonable to assume $L_{x} \gg l$, (where $l\equiv \sqrt{\frac
{\hbar c}{eB}}$ is the magnetic length), and consider a system with
a single edge at $x=0$, where the electron is confined on the half
plane
$-\infty < y < \infty, \ \ \ -\infty < x \le 0$.
Within the Landau gauge ${\bf A}=(0,-Bx)$,
 the Schr\"odinger equation and the corresponding boundary conditions read
 \begin{subequations}
 \begin{eqnarray}
&& \frac{1}{2m}[p_{x}^{2}+(p_{y}+\frac{e}{c} B x)^{2}]
\Psi(x,y)=E\Psi(x,y),
\label{SE2D} \\
&& \Psi(0,y)=\Psi(-\infty,y)=0.
\end{eqnarray}
\end{subequations}
Translation invariance along $y$ enables the replacement $p_{y} \to
k$  and turns equation (\ref{SE2D}) into that of a one dimensional
oscillator centered at
$X_c \equiv k l^{2}$ with hard wall boundary condition at $x=0$.
Using  $l$ as unit of length,
 and $\hbar \omega_c /2 \equiv \hbar^2 /2 m l^{2}$ as unit of energy, the corresponding eigenvalue
problem then reads ($-\infty < x \le 0$)~:
\begin{subequations}
\begin{eqnarray}
&& [-\frac {d^{2}}{dx^{2}}+(x_{c}-x)^{2} ] \psi(x)=E \psi(x),     \label{SE} \\
&& \psi(0)= \psi(-\infty)=0, \label{BCxy}
\end{eqnarray}
\end{subequations}
where all quantities are dimensionless and $-\infty < x_{c}=
{X_c}/{l}=k l< \infty$ is the (dimensionless) center of the
oscillator, which is considered as a continuous parameter. In the
absence of edge,  the eigenvalues are simply the Landau energies
$E_{n}=2n+1$, independent of $x_{c}$. The presence of edge changes
this infinite degeneracy and turns the energies to be dependent on
$x_{c}$. When $|x_{c}|$ is small,
the oscillator center is close to the edge within a magnetic length
so that
 the corresponding eigenfunctions are localized near the edge of the
system at $x=0$ and hence they are
 referred to as {\em edge states}.

The challenge of calculating the eigenstates $\psi_{n}(x;x_{c})$ and
the eigenvalues $E_{n}(x_{c})$ is referred to as (partially)
restricted harmonic oscillator problem, and dates back several
decades ago \cite{Aquino}. Since confined quantum mechanical systems
are ubiquitous, it arises in numerous contexts in physics and
chemistry, much beyond the specific edge-state scenario of the
quantum Hall effect mentioned above. Formally, the eigenvalue
problem (\ref{SE},\ref{BCxy}) possesses an exact solution in terms
of Kummer functions \cite{AS}, but
this solution is hardly useful, as these functions are rather
complicated.
\medskip

 The present work focuses on the eigenvalues
$E_{n}(x_{c})$ of the Schr\"odinger equation (\ref{SE}), analyzing
their dependence on $x_{c}$ and the energy quantum number $n$. It
has two main objectives. The first one is to examine the
semiclassical approach for  estimating the eigenvalues
$E_{n}(x_{c})$.  The idea of employing the WKB approximation for
the study of confined systems has a long history. Probably, the
discussion  most relevant for the present study is a four decade
old work by Vawter \cite{Vawter},  but see also references \cite
{Yakovlev,Sinha,Larsen,Arteca,Isihara,Campoy,Ghatak}. Here the WKB
method is augmented and adapted to the special problem at hand.
The main subtleties resulting from the fact that there are three
regions in $(x_{c},E)$ space (where the WKB connection formulae
assume different forms) are clarified, and the crossover between
these regions is treated with special care.  The energy
eigenvalues are found from an expression $S(E,x_{c})=\pi
[n+\gamma(E, x_{c})]$ relating the WKB action $S(E,x_{c})$ at
energy $E$ with a function $\gamma(E, x_{c})$  which encodes all
the physics behind the connection formulae. The main achievement
of this part is an elucidation of $\gamma(E,x_{c})$ for the entire
range $E>0, \ \ -\infty < x_{c} < \infty$ and a subsequent
evaluation of the eigenvalues $E_{n}(x_{c})$ which appear to be in
excellent agreement with the exact ones.
 As a byproduct,  the source
of disagreement (as noticed in Ref. \cite{Vawter}) between the WKB
solutions and the exact energies occurring when the right turning
point is close to
 the edge, is remedied.

 Our second goal is to examine whether a scaling relation of the form
$E_{n}/L_{n}^{\alpha}=f(x_{c}/L_{n}^{\alpha/2})$,
exists within a wide domain of energies $E_{n}$ and oscillator
center parameter $x_{c}$. Here $L_{n}$ is a linear function of the
energy quantum number $n=0,1,2,\dots$ which serves as a length scale
in energy space, while $\alpha$ is a scaling exponent. Our main
result in this context is that there is indeed an approximate
scaling law (albeit with small correction), which takes the form
\begin{eqnarray}
&&
\frac{E_{n}(x_{c})}{4n+3}=f(\frac{x_{c}}{\sqrt{4n+3}})[1-\frac{3-4
\gamma_{n}(x_{c})}{4n+3}]. \label{scalingfg}
\end{eqnarray}
The function $f(y)$ is a solution of a simple implicit equation ,
(the scaling variable $y \equiv x_{c}/\sqrt{4n+3}$ should not be
confused with the
 $y$ coordinate of the particle).
  The scaling relation (\ref{scalingfg})
of edge state energies is supported by numerical results based on
exact diagonalization.

The rest of the paper is structured as follows: In section
\ref{SecII} the principles of the WKB method as employed here are
recalled in terms of the WKB action and the function
$\gamma(E,x_{c})$. It becomes evident that the magnitude and sign
of $x_{c}$ as well as those of the right turning point crucially
determine the way of constructing the connection formula. The
relevant algorithm is worked out and and the WKB approximated
energies $E_{n}(x_c)$ are obtained and compared with the exact
ones in section \ref{SecIII}. Finally, in section \ref{SecIV} the
scaling hypothesis is tested and a scaling relation for
$E_{n}(x_c)$ is suggested.

\section{Semiclassical quantization
} \label{SecII}

\begin{figure}[!h]
\centering
\includegraphics[width=9.0truecm]{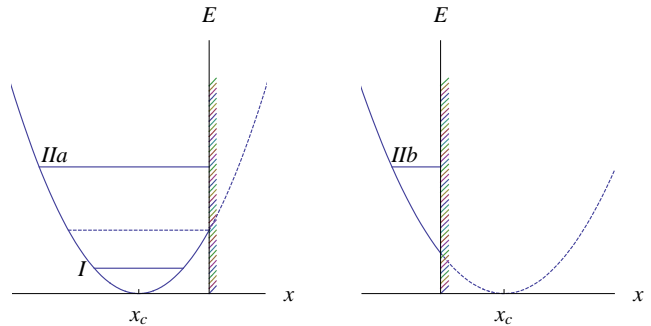}
\caption{\footnotesize Harmonic oscillator potential $(x-x_{c})^{2}$
and a confining wall (edge) at $x=0$. Several regions in $(E,x_{c})$
parameter space
 have to be considered. For
$x_c <0$,  the low energy region ($I$) is defined by $E <x_c^2$, and
the high energy region ($IIa$) is determined by $E
>x_c^2$. For $x_c >0$, there is only a high energy region $IIb$ for which $E
>x_c^2$. The crossover region is marked by the horizontal dashed line in the left panel.}
\label{regimesfig}
\end{figure}

This section briefly introduces and discusses the basic ingredients
of semiclassical action and the function $\gamma(E,x_{c})$ which are
necessary for the subsequent implementation of the WKB analysis.
  For notational convenience the harmonic term $(x-x_{c})^{2}$
  with the hard-wall condition at $x=0$ are
   combined into a single potential,
    \begin{eqnarray}
  && v(x)=\begin{array}{l} \ \ (x-x_{c})^{2}, \ \ \ x<0 \\
 \ \ \  \infty \ \  \ , \ \ \ \ \ \ \ \  x>0 \end{array}
  \end{eqnarray}
    In the WKB method applied for a bound-state problems, the energy $E$ is higher than the potential energy $v(x)$ between the two classical turning points $x_{1 }< x_{2}$ beyond which the solution is classically forbidden.  A glance
    at figure \ref{regimesfig} indicates that the turning points are,
    \begin{eqnarray}
    && x_{1}=x_{c}-\sqrt{E}, \ \ x_{2}=\mbox{min} (x_{c}+\sqrt{E},0),
    \label{tpoints}
    \end{eqnarray}

 The following definitions and notations will be used below in discussing some general features,
 \begin{subequations}
\begin{eqnarray}
&& k(x)=\sqrt{E-v(x)}>0, \ \ \ (x_{1} \le x \le x_{2}), \label{kx}  \\
&& \S(x,x') \equiv \int _{x}^{x'} k(t) d t, \ \ \ (x_{1} \le x \le x' \le x_{2} \le 0), \label{Sxxp} \\
&& S(E,x_{c})=\S(x_{1},x_{2}). \label{S}
\end{eqnarray}
\end{subequations}

The dependence of the WKB action $S(E,x_{c})$ on energy enters
through that of the turning points as in equation (\ref{tpoints}).
This action should satisfy a certain relation resulting from the WKB
connection formulae at the turning points $x_{1}<x_{2}$. In most
cases, the structure of this relation involves the action
$S(E,x_{c})$ as an argument of a trigonometric function, and its
decoding can then be casted in the general form,
\begin{eqnarray}
&& S(E,x_{c})= \pi [n + \gamma (E,x_{c})]. \label{scquant}
\end{eqnarray}
It should be emphasized that equation (\ref{scquant}) is not an
identity, but, rather, an implicit equation for finding the
eigenvalues $E_{n}(x_{c})$. The function $0 \le \gamma (E,x_{c}) <1$
should be calculated independently and encodes the details of the
connection formulae.
 Arriving at this equation (equivalently knowing $\gamma (E,x_{c})$ explicitly) is the central part of the WKB method (the computation of the action $S(E,x_{c})$ is straightforward in most cases). Once this task is achieved, the solutions $E_{n}(x_{c})$ of this implicit equation are the required eigenvalues within the WKB approximation. Inserting the solutions
 into equation (\ref{scquant}) implies the {\em identity}
 \begin{eqnarray}
&& S(E_{n}(x_{c}),x_{c})= \pi [n + \gamma (E_{n}(x_{c}),x_{c})],
 \label{scident}
 \end{eqnarray}
which defines  $\gamma (E_{n}(x_{c}),x_{c}) \equiv
\gamma_{n}(x_{c})$ self consistently as function of the discrete
energy quantum number $n$ and the continuous variable $x_{c}$.
\medskip

The easier part of the WKB procedure is calculation of the action
itself, which in the present problem of confined harmonic oscillator
becomes elementary. It depends on the appropriate region in
$(E,x_{c})$ parameter space, as depicted on figure \ref{regimesfig}.
In the low energy region, $I$ the positions of the turning points
are $x_1=x_c-\sqrt{E}$ and $x_2=x_c+\sqrt{E}$ and the action $S$ is
given by
\begin{eqnarray}
 && S= \int_{x_c-\sqrt{E}}^{x_c+\sqrt{E}} \sqrt{E - (x-x_c)^2} dx= {1 \over 2}
 \pi E . \label{actionI}
 \end{eqnarray}

\begin{figure}[!h]
\centering
\includegraphics[width=9.0truecm]{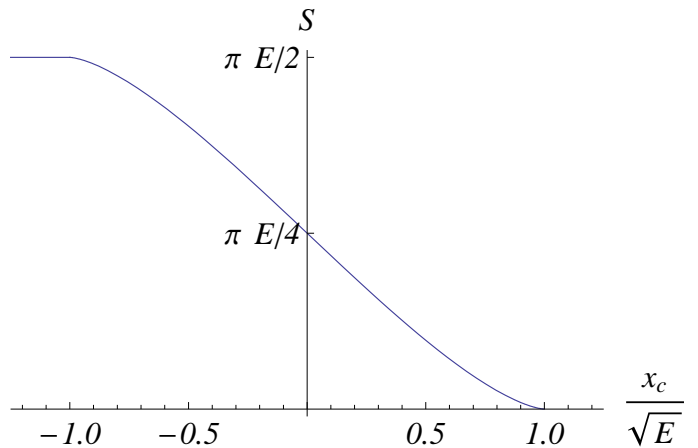}
\caption{\footnotesize The action $S(E,x_c)=\frac{\pi E}{2}s(t)$ as
a function of $t=x_c/\sqrt{E}$ (with $E$ fixed). It varies between
$\pi E/2$ for $t=-1 \ \ (x_c \leq -\sqrt{E})$ and $0$ for $t=1 \ \
(x_c \geq 0)$. For $t=x_c=0$, it is equal to $\pi E/4$. }
\label{actionfig}
\end{figure}

 In both  high energy region $IIa,IIb$, the turning points are located at
  $x_1=x_c-\sqrt{E}$ and $x_2=0$. The
 action is now dependent on $x_c$ and it is given by

\begin{eqnarray}
  S &=& \int_{x_c-\sqrt{E}}^{0} \sqrt{E - (x-x_c)^2} dx \nonumber \\
  &=& {\pi E \over 4} -
{E \over 2}
 \arcsin {x_c \over \sqrt{E}} - {x_c \over 2} \sqrt{E - x_c^2}  .               \label{actionII}
 \end{eqnarray}
 The action has a dimension of energy and can be written as
\begin{eqnarray}
  && S(E,x_c)= {\pi E \over 2} s(x_c/\sqrt{E}) ,         \label{actionIIred}
 \end{eqnarray}
where the dimensionless function
\begin{eqnarray}
  && s(t)=  {1 \over 2}  -
{1 \over \pi}
 \arcsin t - {t \over \pi } \sqrt{1 - t^2 }
            \label{smallaction}
 \end{eqnarray}
 has the following expansions
\begin{subequations}
 \begin{eqnarray}
 t \simeq \pm 1 \qquad && s(t)= {1 \mp 1 \over 2}  \pm  { 4  \sqrt{2} \over 3
  \pi}(1 - |t|)^{3/2}  \\
 t \simeq 0  \qquad && s(t)=  {1 \over 2} - {2 t \over \pi}.
            \label{smallactionexpand}
 \end{eqnarray}
 \end{subequations}
The dependence of the action $S(E,x_c)=\frac{\pi E}{2}s(t)$ on
$t=x_c/\sqrt{E}$ is displayed in figure (\ref{actionfig}).

 The less trivial task which will occupy us in the next subsection is
 to calculate the function $\gamma(E,x_{c})$. When the potential is smooth and the wall is far from the right turning point $x_{2}$, $\gamma$ is determined by the standard WKB procedure such that the solutions for $x<x_{1}$ and $x>x_{2}$ (the classically forbidden regions) contain only decaying exponents. Each turning point $x_{i}$ then contributes a term $\frac {\beta_{i}}{4}$ where $\beta_{i}$ is an {\em integer} (also referred to as the Maslov index for the pertinent turning point), and
 $\gamma=\frac{1}{4}(\beta_{1}+\beta_{2})$. This is the situation appropriate in the lower part of region $I$ in figure \ref{regimesfig} where $\beta_{1}=\beta_{2}=1$, implying $\gamma=\frac{1}{2}$. The definition can be extended to the situation when a turning point occurs at the wall. Although there is no solution to match on the right of the turning point,  the condition $\psi(0)=0$ replaces the connection formula and implies $\beta_{\mbox{wall}}=2$. This is the situation relevant for the higher part of region $IIa$ and for region $IIb$ in figure \ref{regimesfig}, implying $\gamma=\frac{3}{4}$.  As will be stressed below, the situation is different in the crossover region (around the horizontal dashed line in figure \ref{regimesfig}). Since
 $\gamma(E,x_{c})$ is expected to vary smoothly between $\frac{1}{2}$ and $\frac{3}{4}$,
  it cannot be written in terms of an integer $\beta_{2}$. If one insists
  on the same parametrization $\gamma=\frac{1}{4}(\beta_{1}+\beta_{2})$, this implies that $\beta_{2}$ is a non-integer Maslov index \cite{Friedrich}. An attempt to use $\gamma=\frac{1}{2}$ for $E \le x_{c}^{2}$ and $\gamma=\frac{3}{4}$ for $E \ge x_{c}^{2}$
 is too na\"ive and leads to an artificial discontinuity at $E=x_{c}^{2}$ since
 \begin{subequations}
\begin{eqnarray}
&& S(E=x_{c}^{2}+0,x_{c})=\frac {\pi E}{2}=\pi (n+\frac{3}{4})
\nonumber \\
&&\longrightarrow
 E_{n}(x_{c})=2n+\frac{3}{2}, \\
&& S(E=x_{c}^{2}-0,x_{c})=\frac {\pi E}{2}=\pi (n+\frac{3}{4})
\nonumber \\ &&\longrightarrow E_{n}(x_{c})=2n+1.
\end{eqnarray}
\label{disc}
\end{subequations}
This artificial discontinuity is displayed in figure \ref{Enappfig}
below.
\begin{figure}[!h]
\centering
\includegraphics[width=9.0truecm]{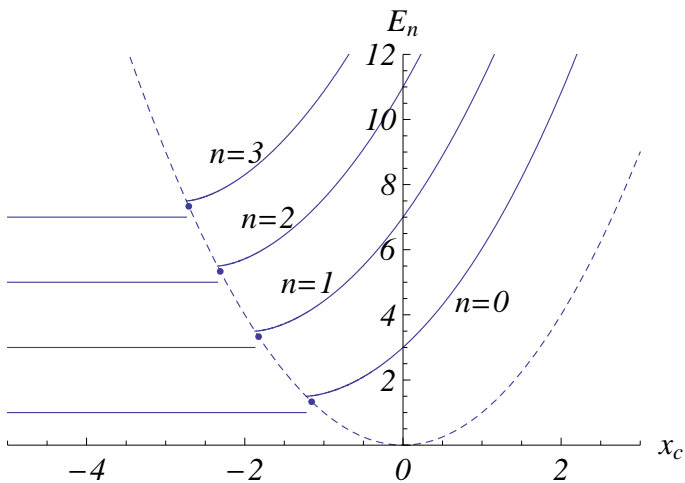}
\caption{\footnotesize  Energy eigenvalues $E_n(x_c)$ obtained
within the semiclassical approximation (\ref{scquant}), where
$S(E,x_{c})$ is given in equations (\ref{actionI},\ref{actionII})
while $\gamma=\frac{1}{2}$ for $E<x_{c}^{2}$ and
$\gamma=\frac{3}{4}$ for $E>x_{c}^{2}$.
 The (artificial) discontinuity occurring along the dashed line $E=x_c^2$ results from an improper treatment of the connection formula  in the crossover region.}
 \label{Enappfig}
\end{figure}
As will be explicitly demonstrated in the next section, the correct
picture is indeed different: for fixed $n$, $\gamma_{n}(x_{c})$ is a
monotonic {\em and smooth} function of $x_{c}$ which tends very
quickly to $\frac{1}{2}$ at negative $x_{c}$ and to $\frac{3}{4}$ at
$x_{c} \ge 0$.
\section{WKB evaluation of eigenvalues}
\label{SecIII} A proper analysis of the crossover regions within the
WKB formalism should then remedy the discontinuity problem by a
proper treatment of the medium energy region $E \simeq x_c^2$. This
is carried below by approaching the cross over region from below ($E
\ll x_{c}^{2} \to E \approx x_{c}^{2}$) and from above ($E \gg
x_{c}^{2} \to E \approx x_{c}^{2}$). The case $x_{c} \approx 0$
where the oscillator center is very close to the wall requires some
special treatment at low energy where the linear approximation for
the potential near the turning point is not valid. After this task
is completed, the eigenvalues calculated by the WKB approach are
compared with the exact ones, and the agreement is virtually
perfect.
\subsection{Approaching $E \approx  x_{c}^{2}$ from below}
\begin{figure}[!h]
\centering
\includegraphics[width=9.0truecm]{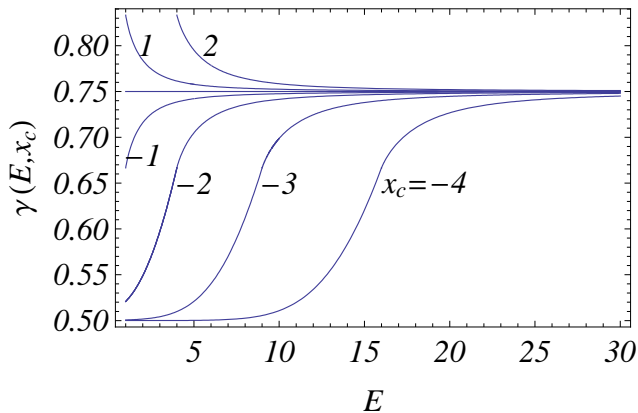}
\caption{\footnotesize The function $\gamma(E,x_c)$ defined in
equation (\ref{gammaI}) is plotted versus $E$ for different values
of  $x_{c}=-4,-3,-2,-1,0,1,2$.  } \label{gammaEfig}
\end{figure}

When  $E \le x_{c}^{2}$ (region $I$ in figure (\ref{regimesfig})),
the turning points and the WKB action are
\begin{subequations}
\begin{eqnarray}
&& x_{1,2}=x_{c} \pm \sqrt{E},  \label{x12} \\
&& S(E)= \frac{1}{2} \pi E. \label{harmonicaction}
\end{eqnarray}
\end{subequations}
 For $E \ll x_{c}^{2}$ the eigenvalues are expected
to approach the  energies $2n+1$ of the unrestricted harmonic
oscillator. The question is how they are modified when $E$
approaches $x_{c}^{2}$ from below. For energies close to $x_c^2$ the
turning point $x_{2}$ is rather close to the wall at $x=0$, and this
must be taken into account. Practically, it means that the wave
function in the classically forbidden region $x_{2}<x<0$ must
include also an exponentially increasing contribution (beside the
exponentially decreasing one), since vanishing of the wave function
at the wall $x=0$ can be achieved only by a proper combination of
the two waves. Therefore, a
 modification of the standard WKB connection procedure is required near $x_{2}$ in
order to take into account the effect of the wall behind the turning
point. On the other hand, the connecting procedure at the left
turning point is standard. Recall that to the left of the point
$x_{1}$ there is a single wave, that decays as $x \to -\infty$. This
means that within the linear approximation for the potential near
the turning points, only the first Airy function $\mbox{Ai}$ is used
near $x_{1}$.  The connecting procedure around $x_{1}$ implies that
the WKB approximation for the wave function  for $x_{1} \le
 x \le x_{2} \le 0$ is
\begin{eqnarray}
  \psi(x)&=&\frac {C}{ \sqrt{k(x)}} \sin[\S(x_{1},x)+\frac
{\pi}{4}] \nonumber \\
&=&\frac {C}{ \sqrt{k(x)}} \sin[S-\S(x,x_{2})+\frac {\pi}{4}],
\label{psiWKB}
\end{eqnarray}
where $C$ is a constant. The modification near the second turning
point $x_{2}$ which is close to the wall assumes that in the small
region $x_{2}<x<0$ the potential is approximated by a linear
expansion near $x_{2}$~:
\begin{eqnarray}
&& v(x) \approx E+2 \sqrt{E}(x-x_{2}). \label{vlinear}
\end{eqnarray}
In terms of the variable
\begin{eqnarray}
&& z=z(x) \equiv (4 E)^{\frac{1}{6}}(x-x_{2}), \label{z}
\end{eqnarray}
and within the approximation (\ref{vlinear}), the Schr\"odinger
equation at the vicinity of $x_{2}$ reads
\begin{eqnarray}
&& \frac{d^{2} \psi(z)}{dz^{2}}-z \psi(z)=0. \label{SEz}
\end{eqnarray}
Unlike the analysis near $x_{1}$, here the combination of {\it both}
Airy functions is required in order to construct the solution around
$x_{2}$, that is,
\begin{eqnarray}
&& \psi(z)=\alpha \mbox{Ai}(z)+\beta \mbox{Bi}(z).  \label{psix2}
\end{eqnarray}
When this combination is examined at $z<0$ (that is, $x<x_{2}$) and
the asymptotic form of the Airy functions is used, it is found that
the wave function at $x < x_{2}$ is,
\begin{eqnarray}
&&\psi(x)=\alpha \sin[\S(x,x_{2})+\frac{\pi}{4}]+\beta \cos
[\S(x,x_{2})+\frac{\pi}{4}]. \label{matchx2}
\end{eqnarray}
Since this is the same function as that found in (\ref{psiWKB}) we
can compare the two and obtain a constraint on the coefficients $C,
\alpha, \beta$. To this end, denote $\zeta \equiv
\S(x,x_{2})+\frac{\pi}{4}$, and open the sine and cosine functions,
to get
\begin{eqnarray}
&& \sin \zeta (C \sin S+ \alpha)+\cos \zeta (C \cos S + \beta)=0.
\label{matching1}
\end{eqnarray}
This holds for every $\zeta$, in the appropriate interval, hence
each term in the bracket should vanish separately. This yields a
couple of equations for the three coefficients $C, \alpha, \beta$.
The wave function (\ref{psix2}) should vanish at the wall $x=0$,
that is ($ z_0 \equiv z(x=0)$)~:
\begin{subequations}
\begin{eqnarray}
&& \psi(z_0)=0, \label{psiwall} \\
&& z_0  = -(4 E)^{\frac{1}{6}}x_{2}  =-(4
E)^{\frac{1}{6}}(x_{c}+\sqrt{E})>0, \label{zL}
\end{eqnarray}
\end{subequations}
implying that
\begin{eqnarray}
&& \beta=-\frac{\mbox{Ai}(z_0)}{\mbox{Bi}(z_0)} \alpha.
\label{betalpha}
\end{eqnarray}
Employing the statement after equation (\ref{matchx2}) this leaves
us with two homogeneous equations for $C$ and $\alpha$~:
\begin{subequations}
\begin{eqnarray}
&& C \sin S + \alpha = 0, \label{eq1} \\
&& C \cos S -\frac{\mbox{Ai}(z_0)}{\mbox{Bi}(z_0)} \alpha=0,
\end{eqnarray}
\label{eq2}
\end{subequations}
and the energies $E_{n}$ are obtained by requiring the vanishing of
the determinant
\begin{eqnarray}
&& \cos S+\frac{\mbox{Ai}(z_0)}{\mbox{Bi}(z_0)} \sin S=0.
\label{det}
\end{eqnarray}
The implicit dependence on the energy $E$ enters through the
dependence of $S$ and $z_0$ on energy, equations
(\ref{harmonicaction}) and (\ref{zL}) respectively. Equation
(\ref{det}) is recast in the WKB format as
\begin{eqnarray}
&&  S= {\pi E \over 2} = \pi  [n + \gamma(E,x_c)]
\label{actiongammaExc}
\end{eqnarray}
where $0 < \gamma <1$ is explicitly given by
\begin{eqnarray}
&&   \gamma(E,x_c)= 1  - {1 \over \pi} \arctan {\mbox{Bi}(z_0) \over
\mbox{Ai}(z_0) }, \label{gammaI}
\end{eqnarray}
with $z_0$ defined in (\ref{zL}). The function  $ \gamma(E,x_c)$ is
plotted as function of the energy $E$ for several values of $x_c$ in
figure \ref{gammaEfig}.
The eigenvalues $E_{n}(x_{c})$ are the intersection of the curves
$\gamma(E,x_{c})$ (as function of $E$ with $x_c$ as a parameter)
with the straight lines $\frac{1}{\pi}S(E,x_{c})-n$. In the two
limiting cases $E \ll x_{c}^{2}$ and $E \approx x_{c}^{2}$ these
solutions can be easily evaluated. When $E \ll x_c^2$, the value of
$z_0$ is large. Then
\begin{eqnarray}
&& \frac {\mbox{Ai}(z_0)}{\mbox{Bi}(z_0)} \approx
\frac{1}{2}e^{-\frac{4}{3}z_0^{\frac{3}{2}}} \approx 0,
\label{ratio1}
\end{eqnarray}
and equation (\ref{gammaI})  leads to the familiar result of the
harmonic oscillator problem with $\gamma=1/2$ and $E_{n}=2n+1$. On
the other hand, when the energy is very close to (but still lower
than) $x_c^2$ then
 $z_0$ is very small, and hence,
\begin{eqnarray}
&& \frac {\mbox{Ai}(z_0)}{\mbox{Bi}(z_0)}
 \approx  \frac {\mbox{Ai}(0)}{\mbox{Bi}(0)} = \frac{1}{\sqrt{3}}, \label{ratio11}
\end{eqnarray}
so that $\gamma=2/3$ and
\begin{eqnarray}
&& E_{n}=2n+\frac{4}{3}. \label{mediumenergy}
\end{eqnarray}
This value $\gamma=\frac{2}{3}$ is intermediate between the "high"
and "low" energy values $\frac {3}{4}$ and $\frac {1}{2}$. For $n=0$
this gives the energy $\frac{4}{3}$ instead of $\frac{3}{2}$ and
$1$. The exact value is indeed $\frac{4}{3}$. The disagreement
between the "high energy" prediction $\frac{3}{2}$ and the exact
value $\frac{4}{3}$ has been noticed for $n=0$ in table II of
Ref.\cite{Vawter} where it is dubbed as a WKB error.  Our results
show, however, that the WKB formalism as applied here is virtually
exact.


\subsection{Approaching $E \approx  x_{c}^{2}$ from above}
Consider now region (II) in figure \ref{regimesfig}, where $E \ge
x_{c}^{2}$.
 In order
to treat this "high energy" region the linear expansion of $v(x)$,
eq. (\ref{vlinear}) is now replaced by
\begin{eqnarray}
&& v(x) \approx x_{c}^{2} - 2 x_{c} x ,\label{vlinear1}
\end{eqnarray}
so that the wave function   satisfies  the Schr\"odinger equation
and the hard wall boundary conditions at $x=0$,
\begin{subequations}
\begin{eqnarray}
&& \psi'' - (x_c^2 - E -2 x x_c) \psi =0. \label{SEHE} \\
&& \psi(0)=0. \label{HWBC}
\end{eqnarray}
\end{subequations}
whose solution has the form
\begin{eqnarray} \psi(x)= \alpha \mbox{Ai} \left[{x_c^2 - E -2 x x_c \over
|2x_c|^{2/3}}\right] + \beta \mbox{Bi} \left[ {x_c^2 - E -2 x x_c
\over |2x_c|^{2/3}}\right] . \label{psinew} \end{eqnarray}
The hard wall condition at $x=0$ implies

\begin{eqnarray}  {\beta \over \alpha}= - \displaystyle {\mbox{Ai} \left[{x_c^2 -
E  \over |2 x_c|^{2/3}}\right] \over \mbox{Bi} \left[{x_c^2 - E
\over |2 x_c|^{2/3}}\right]} \equiv - {\mbox{Ai} (-W) \over
\mbox{Bi} (-W)} , \label{betalpha1} \\ \end{eqnarray}
where
\begin{eqnarray}  W= {E - x_c^2 \over |2 x_c|^{2/3}} >0 .\label{defW}
\end{eqnarray}

 At  large distance $x <0$,  the asymptotic form of the  Airy functions for negative arguments
 can be used to yield
\begin{eqnarray}  \psi(x) &\propto& \alpha \sin \left[{x_c^2 - E -2 x x_c \over
|2x_c|^{2/3}} + {\pi \over 4}\right] \nonumber \\ &+& \beta \cos
\left[{x_c^2 - E -2 x x_c \over |2x_c|^{2/3}}+{\pi \over 4}\right] .
\label{psinew2}
\end{eqnarray}
and the action for the linearized potential $v(x)$ is
\begin{eqnarray} \S(x,0)&=&\int_x^0 \sqrt{E - x_c^2 + 2 x x_c} dx \nonumber \\
&=&   {(E - x_c^2 )^{3/2} \over 3 x_c} -{(E - x_c^2 + 2 x x_c)^{3/2}
\over 3 x_c}  .          \label{SAiry1}
\end{eqnarray}
Therefore the wavefunction has the form

\begin{eqnarray}   \psi(x) &\propto& \alpha \sin\left[ - \epsilon \S(x,0) + \delta + {\pi \over
4}\right] \nonumber \\ &+& \beta \cos\left[ -\epsilon \S(x,0) +
\delta+ {\pi \over 4}\right] , \label{psinew3}
\end{eqnarray}
with
\begin{subequations}
\begin{eqnarray}
&& \epsilon= \mbox{sign}(x_c), \\
&& \delta= {(E - x_c^2)^{3/2} \over 3 |x_c|}={2 \over 3} W^{3/2}.
\label{epsdelta}
\end{eqnarray}
\end{subequations}
On the other hand, following the procedure leading to equation
(\ref{psiWKB}) the wave function resulting from the connecting
formula at $x=x_{1}$ is ($x_{1} \le x \le 0$)
\begin{eqnarray}
  \psi(x)&=&\frac {C}{ \sqrt{k(x)}} \sin[S-\S(x,0)+\frac
{\pi}{4}] \nonumber \\ &=& \frac {C}{ \sqrt{k(x)}}
\cos[\S(x,0)+\frac {\pi}{4}-S]  . \label{psiWKB1}
\end{eqnarray}
 It is now allowed  to
compare the two expressions   (\ref{psinew3}) and (\ref{psiWKB1})
for the same wave function. These two expressions should be equal
  for any value of the variable $\S(x,0)$.
Equating the coefficients of $\sin \S(x,0)$ and $\cos \S(x,0)$  to
$0$ (each one separately), and employing the relation
(\ref{betalpha1}) then result in the following two equations for $C$
and $\alpha$~:
\begin{subequations}
\begin{eqnarray}
 -\alpha \epsilon \cos (\delta + {\pi \over 4} ) + \beta \epsilon
\sin (\delta+{\pi \over 4})
&=&  -C  \cos (S+{\pi \over 4}),  \nonumber \\
  \alpha  \sin (\delta+{\pi \over 4}) + \beta \cos (\delta+{\pi
\over 4}) &=& C \sin (S+{\pi \over 4}),  \label{Aalpha1}
\end{eqnarray}
\end{subequations}
leading to
\begin{eqnarray} && \tan ( S +{\pi\over 4})= \epsilon \cot (\delta_0 - \delta +{\pi\over 4})   \end{eqnarray}
\ \\
with
\begin{eqnarray} && \tan \delta_0= {\alpha \over \beta}= - {\mbox{Bi} (-W) \over
\mbox{Ai} (-W)}.
\end{eqnarray}

Finally, the energy levels $E(x_c)$ are given by the action
(\ref{actionII}) and

\begin{eqnarray} && \tan (S+{\pi \over 4})= -\epsilon  \cot \left(  \arctan {\mbox{Bi}[ -W] \over
\mbox{Ai}[-W]} +{2 \over 3} W^{3/2}+ {\pi \over 4} \right) \nonumber
\\ &&
\end{eqnarray}
\ \\
or, equivalently, by the central expression,
\begin{eqnarray} && S(E,x_c)=  \pi [n + \gamma(E,x_c)] \end{eqnarray}
\ \\
where $0< \gamma(E,x_c)<1$ is given explicitly by (compare with
equation (\ref{gammaI}) applicable for the case $E \le x_{c}^{2}$),
\begin{eqnarray}
&& \gamma(E,x_c) =  {3 \over 4}  \nonumber \\
&&- {\epsilon \over \pi}
 \arctan \cot \left(  \arctan \frac{\mbox {Bi}}{\mbox{Ai}} +\frac {2}{3}W^{\frac{3}{2}} + \frac{\pi}{4} \right)
 \label{gammaexp2}
\end{eqnarray}
\ \\
This function, calculated here for $x_c > -\sqrt{E}$ is shown on
figure (\ref{gammaEfig}). It has simple values in the following
special cases~:
\begin{subequations}
\begin{equation}
 x_c = -\sqrt{E}, \quad W=0, \quad \gamma(E,x_c)=2/3, \quad E_n= 2
n +{4 \over 3}, \label{en43}
\end{equation}
\begin{equation}
 x_c =0, \quad W=\infty, \quad
\gamma(E,x_c) = 3/4,  \quad E_n= 4 n +3.  \label{en4np3}
\end{equation}
\end{subequations}

\subsection{{ Perturbation expansion for small $|x_c|$}}
Near $x_{c}=0$ and at low energy the linear approximation is not
justified because at $x_{c}=0$ the linear term vanishes. It is then
necessary to modify the semiclassical formalism for evaluating
$E_n(x_c)$ and $\gamma_n(x_c)$ near $x_c=0$. Here  a simple
perturbative expansion for small $x_{c}$ is used. The unperturbed
hamiltonian is,

\begin{eqnarray}
&& H_0= p^2 + x^2
\end{eqnarray}
The solutions $\psi_{n}(x)$ with $\psi(0)=0$ are  the antisymmetric
eigenfunctions of the free harmonic oscillator (with energy
$E_{n}=4n+3$),
\begin{eqnarray}
&&  \psi_n(x)={e^{-x^2/2} H_{2 n+1}(x) \over \sqrt{\pi 2^{2 n } (2
n+1)!}}. \label{inpair}
\end{eqnarray}
where $H_{n}$ are the Hermite polinomyals. The perturbation term for
small $x_{c}$  is $- 2 x  x_c$, and the first order correction then
yields the perturbed energies,
\begin{eqnarray}
    E_n(x_c) &\approx& 4 n +3 - 2 x_c \int_{- \infty}^0 x
\psi_{n}^2(x)dx \nonumber \\
&=&(4 n +3) + {4 x_c \over \sqrt{\pi}} \displaystyle \prod_1^n
\left( 1 + {1 \over 2 p} \right) \label{perturb}
\end{eqnarray}
The value of $\gamma_c(x_c)$ near $x_{c}=0$ can therefore be
obtained from the quantization condition eq. (\ref{scident}).
Expanding the action $S(E,x_c)$ near $x_c=0$ to first order in
$x_{c}$ reads  $S(E,x_c)=\pi E/4 - \sqrt{E} x_c$,  which implies,
\begin{eqnarray}
&&   \gamma_n(x_c)= {3 \over 4}  + \beta_n x_c. \label{gammaperturb}
\end{eqnarray}
The coefficients $\beta_{n}$ are given by,
\begin{eqnarray} \beta_n &=& {1 \over \sqrt{\pi}}  \displaystyle \prod_1^n \left( 1 + {1
\over 2 p} \right) - {\sqrt{4 n +3} \over \pi} \nonumber \\
\beta_0 &=& 0.01286 \nonumber \\
\beta_1 &=& 0.00416 \nonumber \\
\beta_2 &=& 0.00214 \nonumber \\
\beta_3 &=& 0.00135 \end{eqnarray} In the limit $n \rightarrow
\infty$, the   product tends to $\sqrt{{4 n + 3 \over \pi}}$, so
that $\beta_\infty=0$. Equation (\ref{gammaperturb}) shows that at
low energy and for $x_{c}$ very close to $0$ (when the linear
approximation to the potential fails), $\gamma_{n}(x_{c})$ shoots up
slightly above $\frac{3}{4}$.
Since $\beta_{n}$ is small, the deviation $\beta_{n} x_{c}$ is
virtually negligible.

\subsection{Comparing WKB results with the exact ones}
\begin{figure}[!h]
\centering
\includegraphics[width=9.0truecm]{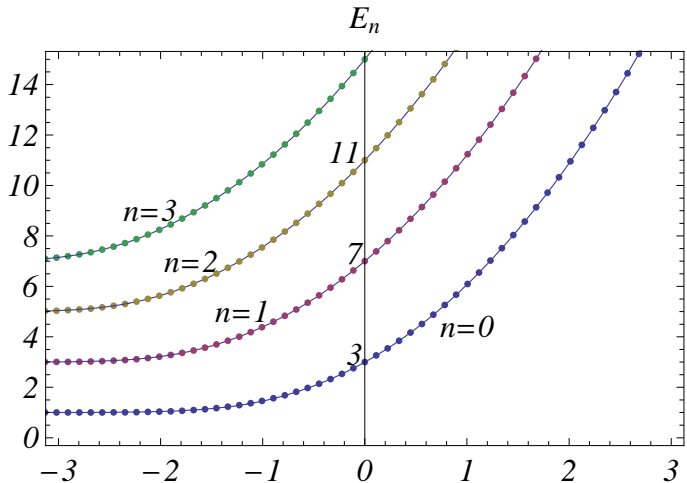}
\caption{\footnotesize  The solid lines display the energy levels
$E_n(x_c)$ calculated in the WKB approximation as function of the
continuous parameter $x_c$ . The exact eigenvalues (obtained by
numerical diagonalization) are represented by the full circles.}
\label{Engoodfig}
\end{figure}

The central result of the foregoing discussion can now be presented.
Once the function $\gamma(E,x_c)$ is known, the spectrum $E_n(x_c)$
is
 entirely determined by the implicit equation (\ref{scquant} ). This is calculated
 and displayed as solid lines in figure (\ref{Engoodfig}). The exact eigenvalues
 are marked by dots on the same figure. The fit is
indeed perfect.
\\


\section{Inspecting scaling behavior}
\label{SecIV}

After substantiating the efficiency   of the WKB method for the
confined harmonic oscillator problem (summarized in figure
\ref{Engoodfig}), and  elucidating the functional form of
$\gamma(E,x_{c})$ (see equations (\ref{gammaI},\ref{gammaexp2})) let
us return to the question of scaling posed in the Introduction. It
should be stressed that the notion of scaling here simply means that
some functions of $x_{c}$ and $n$ can be represented as functions of
a certain combination of these two variables. It has nothing to do
with thermodynamics, of course, although the variable $n$ is, in
some sense, analogous to the notion of length scale.
\subsection{Scaling of $\gamma_{n}(x_c)$}
Before discussing the scaling of the energies themselves it is
useful to check a possible scaling behavior of $\gamma_{n}(x_c)$.
\begin{figure}[!h]
\centering
\includegraphics[width=9.0truecm]{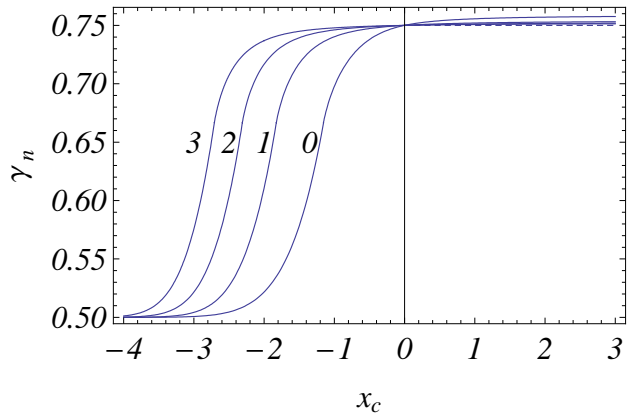}
\caption{\footnotesize The function $\gamma_n(x_c)$ as function of
$x_c$ for $n=0,1,2,3$.  Its calculation is explained at the
beginning of this subsection.} \label{gammascfig}
\end{figure}
To obtain this function explicitly the eigenvalues $E_n(x_c)$ are
calculated as explained in the previous section, and then
substituted instead of $E$ in the function $\gamma(E,x_{c})$ defined
in equations (\ref{gammaI},\ref{gammaexp2}))  leading to the desired
function $\gamma_n(x_c)=\gamma[E_n(x_c),x_c]$. It is displayed as
function of $x_{c}$ for several values of $n$ in
 figure (\ref{gammascfig}). Looking at figure \ref{gammascfig} one is tempted to search for a
"scaling relation" in the sense that  $\gamma_{n}(x_c)$ depends on a
certain combination $X(n,x_{c})$ of $x_{c}$ and $n$.  Such
combination can then be used as a scaling variable where all the
curves collapse on a single one. Based on our previous analysis it
is established that $\gamma_n(x_c)$ and the spectrum $E_n(x_c)$ is
very well fitted (within 2 \%) by the function

\begin{eqnarray}
 \gamma_n(x_c)&=& 1    \\
  &-& {1 \over \pi} \arctan {\mbox{Bi}[-2^{1/3}(2
n +{ 4 \over 3})^{1/6} (x_c + \sqrt{2n +{4 \over 3}})] \over
\mbox{Ai}[-2^{1/3}(2 n +{ 4 \over 3})^{1/6} (x_c + \sqrt{2n +{4
\over 3}})] } \nonumber  \label{gammafitI}
\end{eqnarray}
for $x_c < - \sqrt{2 n + {4 \over 3}}$, and by the function

\begin{eqnarray} && \gamma_n(x_c)=  1- {1 \over \pi} \arctan \tan \left(
\arctan {\mbox{Bi}[ -W_n] \over \mbox{Ai}[-W_n]} +{2 \over 3}
W_n^{3/2} \right) , \nonumber \\
&& \mbox{with} \qquad W_n(x_c)= {2 n +{4 \over 3} - x_c^2  |2
x_c|^{2/3}},
\end{eqnarray}
for $ - \sqrt{2 n + {4 \over 3}}<x_c <0$ and by $\gamma_n(x_c)=3/4$
for $x_c <0$.
\bigskip

\begin{figure}[!h]
\centering
\includegraphics[width=9.0truecm]{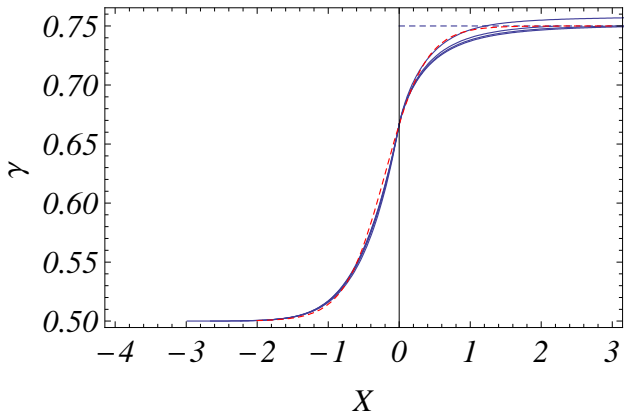}
\caption{\footnotesize  The functions $\gamma_{n}(x_{c})$ are
plotted as a function of the scaling variable $X= (2n +{4 \over
3})^{1/6} (x_c +\sqrt{2 n +{4 \over 3}})$. The dashed line is the
function $\gamma(X)$ defined by equation (\ref{gammX}).}
\label{scalgamm}
\end{figure}

Strictly speaking then, there is no single combination of $n$ and
$x_{c}$ that enters both expressions. Practically, however, when the
variable $X= (2n +{4 \over 3})^{1/6} (x_c +\sqrt{2 n +{4 \over 3}})$
is used, the collapse of all curves is very good as can be seen in
figure \ref{scalgamm} where $\gamma_n(x_c)= \gamma(X)$,  is plotted
against $X$.

Moreover, $\gamma(X)$ is very accurately fitted by the function
\begin{eqnarray}
&& \gamma(X)= {3 \over 4} - {1 \over 4} {1 \over 2 \exp A X +1}
\label{gammX}
\end{eqnarray}
 where $A \simeq 3.5$, represented by the dashed line in figure \ref{scalgamm}.
The energies $E_{n}(x_c)$
 are then calculable (within 2 \% accuracy) through the solution of the standard WKB implicit equation,
\begin{eqnarray}
&&  S(E_n,x_c)= \pi  [n + \gamma(X)] \label{actiongammaExc2}
\end{eqnarray}
with  $\gamma (X)$ as given by the approximate formula
(\ref{gammX}).

\subsection{Scaling of edge state energies $E_{n}(x_c)$}
Turning now to the possible scaling of edge state energies, consider
the identity (\ref {scident}) and suppose, for the moment that
 $\gamma_{n}(x_{c})$ is a constant. After dividing
 both sides by $2 \pi (n+\gamma)$ and defining the scaling variables
 \begin{subequations}
 \begin{eqnarray}
&&  y=x_c/2 \sqrt{ n + \gamma}, \label{yscal} \\
 && u= E/4 (n+ \gamma), \label{uscal}
 \end{eqnarray}
 \end{subequations}
 the identity (\ref{scident}) takes the form
 \begin{eqnarray}
 && \frac{1-u}{2}+\frac{u}{\pi} \arcsin \frac{y}{\sqrt{u}}+\frac{y}{\pi} \sqrt{u-y^{2}}=0.
 \label{fofy}
  \end{eqnarray}
  Equation (\ref{fofy}) then implicitly defines the functional relation
  \begin{eqnarray}
  && u=\frac{E_{n}}{4(n+\gamma)}=f(y)=f(\frac{x_{c}}{2\sqrt{n+\gamma}}), \label{uf}
  \end{eqnarray}
   between $u$ and $y$. The function $f(y)$ is shown in figure \ref{scalingfig} below.
\begin{figure}[!h]
\centering
\includegraphics[width=9.0truecm]{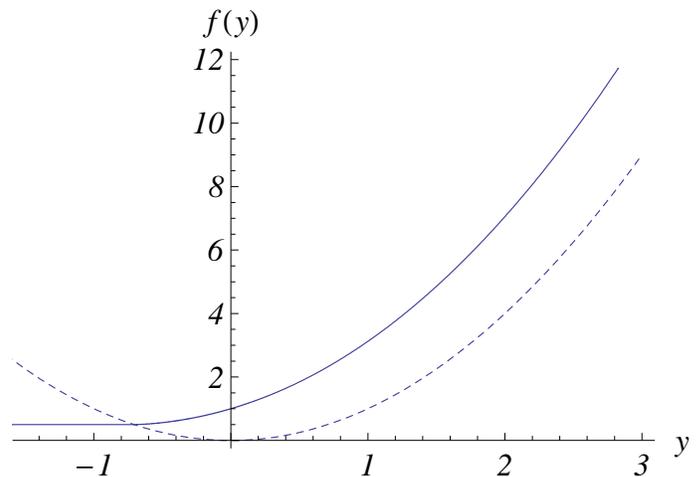}
\caption{\footnotesize  Universal function $f(y)$ and $y^2$ (dashed
line). } \label{scalingfig}
\end{figure}
From the expansions   (\ref{smallactionexpand}) of the function
$s(t)$, it is easy to deduce that the function $f(y)$,  shown on
figure (\ref{scalingfig}), has the following limits
\begin{equation}
 y \leq -1/\sqrt{2}  \ \ f(y)={1\over 2},  \nonumber
 \end{equation}
 \begin{equation}
0> y \geq -1/\sqrt{2}  \  \  f(y)={1\over 2} + {4 \ 2^{1/4}
\over 3 \pi} ({1 \over \sqrt{2}}+y)^{3/2}, \nonumber
\end{equation}
 \begin{equation}
   y \simeq 0  \  \
   f(y)=1+ {4 y \over \pi}, \nonumber
   \end{equation}
 \begin{equation}
 y \rightarrow \infty  \  \ f(y)=y^2 +{1 \over 2} \left({3
\pi \over \sqrt{2}}\right)^{2/3} y^{4/3}. \label{limitsfoy}
 \end{equation}
However, the relation (\ref{fofy}) is of little use, because
$\gamma$ is not known in advance and its evaluation is part of the
problem.
 A more practical approach would then be to fix an appropriate constant value $\gamma_{0}$ and
take account of the small variation of $\frac{1}{2} \le \gamma \le
\frac{3}{4}$ by some correction. From the edge-state point of view,
the region $x_{c}>-1$ for which $\gamma  \approx \frac{3}{4}\equiv
\gamma_{0}$ is the most relevant one. Therefore,  the "length scale"
is fixed as $ L_{n}=4(n+\gamma_{0})=4n+3$ and expand equation
(\ref{uf}) to first order around $\gamma_{0}$. To save notations $u$
and $y$ are now defined with $\gamma=\gamma_{0}=3/4$.  The result is
\begin{eqnarray}
&& \frac{E_{n}}{4n+3}=u=f(y)\left \{1-\frac{3-4 \gamma}{4n+3}\left
[1-\frac{y f'(y)}{2 f(y)} \right ] \right \}.
\label{Enfinal}
\end{eqnarray}
Strictly speaking, the problem mentioned in connection with equation
(\ref{fofy}) still remains, but this time it is casted in a form
showing that
 exact scaling is expected to work for large $n$  as $L_{n}=4n+3 \to \infty$. The second term in the square brackets can be neglected because for large $y$ $\gamma=3/4$ and the correction disappears, whereas for small $y$, $f(y) \to 1/2$ so its derivative vanishes.  It is then reasonable to
  suggest the following relation
 \begin{eqnarray}
 && \frac{E_{n}(x_{c})}{4[n+\gamma_{n}(x_{c})]}=f(y),  \label{scalfinal}
 \end{eqnarray}
 where,  employing the expression (\ref{gammX}) for $\gamma_{n}(x_{c})$,
 \begin{eqnarray}
  && 4[n+\gamma_{n}(x_{c})]=4n+3-\frac{1}{2 e^{A X_{n}(x_{c})}+1}, \label{scalen} \\
 && \mbox{with} \ \ \ X_{n}(x_{c})= (2n +{4 \over 3})^{1/6}
\left [x_c +(2 n +{4\over 3})^{\frac{1}{2}} \right ],
 \nonumber
\end{eqnarray}
with $A= \simeq 3.5$. To test the scaling relation, the exact eigenvalues $E_{n}(x_{c})$ for
$n=0,1, \ldots 13$ were calculated via numerical diagonalization
resulting in
  $14$ curves each contains $100$ points $x_{c}$ (we have already shown that these energies can also be calculated within the WKB formalism developed here).
 \begin{figure}[!h]
\centering
\includegraphics[width=9.0truecm]{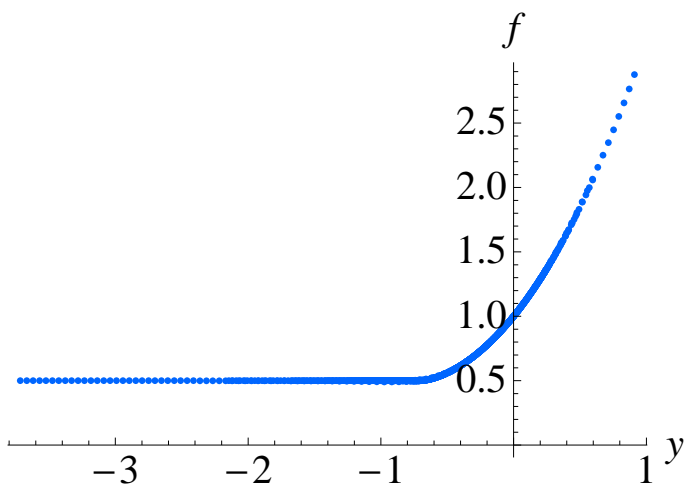}
\caption{\footnotesize Test of the scaling relation (\ref{scalfinal}). The numbers
$E_n(x_c)$ are obtained numerically by exact
diagonalization and the function $\gamma(x_c)$ is approximated
employing equation (\ref{scalen}). The numbers appearing on the LHS of equation
(\ref{scalfinal}) are then displayed as function of the scaling
variable $y=\frac{x_{c}}{\sqrt{4n+3}}$, thus generating a smooth curve representing
the scaling function $f(y)$ given by eqs. (\ref{fofy},\ref{uf}). It
results from the collapse of $14$ curves $E_n(x_c)$ with $n=0,1,\dots 13$,
each contains $100$ points
$x_{c}$. } \label{scalingtest}
\end{figure}
 The $1400$ numbers appearing on the LHS of equation (\ref{scalfinal}) are thus computed and displayed against the scaling variable $y$ in figure \ref{scalingtest} below.
 The smooth curve gives the scaling function $f(y)$ and for $y > -\frac{1}{\sqrt{2}}$ it coincides with the
 curve of figure \ref{scalingfig}.  The collapse of $14$ curves on a single one is remarkable. The upshot is that equation (\ref{scalfinal}) with $f(y)$ analyzed in equations (\ref{limitsfoy}) and $\gamma(x_c)$ parametrized as in equation (\ref{gammX}) is an excellent approximation for $E_{n}(x_c)$.

\noindent
{\bf Acknowledgment} \\
We would like to thank Jean Marc Luck and Benoit Dou\c cot for
invaluable help and suggestions.


\begin{thebibliography}{99}
\bibitem{BH} B. I. Halperin, Phys. Rev. {\bf B} (1982).
\bibitem{MD} A. H. Macdonald and P. Streda, Phys. Rev. {\bf B29}, 1616 (1984).
\bibitem{MB} M. B\"uttiker, Phys. Rev. {\bf B} (1988).
\bibitem{CK} C. Kane and E. J. Mele, Phys. Rev. Lett. (2005).
\bibitem{Aquino} H. E. Montgomery Jr, G. Campoy and N. Aquino, arXiv:0803.4029 (Refs. 17-39 therein), (2008).
\bibitem{AS} M. Abramowitz and Irene A. Stegun, {\em Hanbook of Mathematical Functions}, National Bureau of Standards, Applied Mathematic Series (1964). See Chapter 19, {\em Parabolic Cylinder Functions}.
\bibitem{Vawter} R. Vawter, Phys. Rev. {\bf 174}, 749 (1968).
\bibitem{Yakovlev} D. S. Kr\"ahmer, W. P. Schleich and V. P. Yakovlev, J. Phys. A: Math. Gen. {\bf 31}, 4493 (1998).
\bibitem{Sinha} A. Sinha and R. Ryochudhury, Int. Jour. of Quantum Chemistry, {\bf 73}, Issue 6, 497 (1999).
\bibitem{Larsen} U. Larsen, J. Phys. A: Math. Gen. {\bf 16}, 2137 (1983).
\bibitem{Arteca} G. A. Arteca, S. A. Maluendes, F. M. Fernandez, E. A. Castro, Int. Jour. of Quantum Chemistry, {\bf 24}, Issue 2, 497 (1983).
\bibitem{Isihara} A. Isihara and K. Ebina, J. Phys. C: Solid State Physics, {\bf 21}, L1079 (1988).
\bibitem{Campoy} G. Campoy, N. Aquino and V. D. Granados, J. Phys. A: Math. Gen. {\bf 35}, 4903 (2002).
\bibitem{Ghatak} A. K. Ghatak, I. C. Goyal, R. Jindal and Y. P. Varshni, Can. J. Phys. {\bf 76}(5), 351 (1998).
\bibitem{Friedrich} H. Friedrich and J. Trost, Phys. Rev. {\bf A54}, 1136 (1996).
\end{thebibliography}
\end{document}